\pdfoutput=1

\documentclass[11pt]{article}
\usepackage{geometry}                
\usepackage{amsmath}
\usepackage{vmargin}
\usepackage{graphicx,color}
\usepackage{subcaption}
\usepackage{overpic}
\usepackage{algorithm}
\usepackage{algpseudocode}
\usepackage{multirow}

\DeclareMathOperator*{\argmin}{arg min}

\newcommand{\vc}[1]{\ensuremath{\mathbf{#1}}}
\newcommand{\trp}{^{\text{\scriptsize T}} }

\newcommand{\inv}{^{-1} }
\newcommand{\sbr}[1]{\ensuremath{_{\mathrm{#1}}}}
\newcommand{\spr}[1]{\ensuremath{^{\mathrm{#1}}}}

\begin{document}

\title{Metropolized Randomized Maximum Likelihood \\ for sampling from multimodal distributions}
\author{Dean S. Oliver \\
Uni Research, Uni CIPR \\
Bergen, Norway }

\maketitle

\section*{Abstract}

This article describes a method for using optimization to derive efficient independent transition functions for  Markov chain Monte Carlo (MCMC) simulations with application to sampling from a posterior density $\pi(\vc{x})$  which   is multimodal. Although the proposals in the Randomized Maximum Likelihood method are placed in regions of high probability, the distribution of samples is only approximately correct. Introduction of auxiliary variables simplifies the computation of the proposal density, allowing the Metropolis-Hastings method to be applied. 
We restrict our attention to the special case  for  which the target density is the product of a multivariate Gaussian prior and a  likelihood function for which the errors in observations are additive and Gaussian.

\section*{Introduction}

Many geoscience inverse problems are characterized by nonlinearity in the relationship between parameters and observations \cite{tarantola:82b,carrera:05,oliver:08}.  In some cases, the posterior pdf for the model parameters after assimilation of observations is multimodal, although the presence or absence  of multiple modes in large models is seldom verified. Examples of smaller models in which multiple modes have been verified include flow and transport problems in  porous media, where the nonlinearity resulting from uncertainty in connections in layered media and observation operators that average over layers can both lead to multiple modes \cite{christie:06,oliver:11b}.
Monte Carlo sampling  is often the only viable method of quantifying uncertainty in the posterior distribution for problems of this type, but the cost of obtaining a substantial number of independent samples can be prohibitive. 

In order to obtain a reasonably high probability of generating acceptable random-walk transitions in  Metropolis sampling, the transitions must generally be small: for small enough transition distances, almost all proposals will be accepted. In that case, however, 
it is possible to  remain trapped in some modes for very long times before jumping to another mode and the time required to obtain useful Monte Carlo estimates can be impractical \cite{fort:14}. The problem of optimal scaling of the proposal distribution to maximize the efficiency of mixing of a random walk Metropolis algorithm has been  solved for certain classes of target distributions \cite{roberts:97,roberts:01}. These scaling rules are not applicable in all contexts, and in particular are not appropriate for multimodal distributions \cite{fort:14a}.
The ability to design  a proposal distribution that mixes well is key to the efficiency of the MCMC algorithm for multimodal distributions.

In this paper, we discuss an augmented variable independence Metropolis sampler that uses minimization to place proposals in regions of high probability. 
Augmented variable methods have been shown to be useful  in MCMC to construct efficient sampling algorithms  and in particular for sampling from multimodal distributions \cite{besag:93}, where the auxiliary variables might allow for large jumps between modes \cite{tjelmeland:04}.
Storvik \cite{storvik:11} suggests that auxiliary variables can be used in Metropolis-Hastings (MH) either for generating better proposals or for easier calculation of acceptance probabilities and points out that the use of auxiliary variables allows flexibility in choosing the target distribution in the augmented space, as long as the marginal distribution for the variables of interest is unchanged. 
In this paper, the augmented variables are useful for obtaining a marginal proposal density that is close to the target marginal density, and for simplifying evaluation of the MH ratio. 

Independence Metropolis samplers are Metropolis-Hastings samplers for which the transition probability $q(\vc{y}|\vc{x})$ does not depend on the current state of the chain ($q(\vc{y}|\vc{x}) = q(\vc{y})$) \cite{tierney:94}.
For  independence sampling, it is useful to choose the proposal density $q(\vc{x})$ such that $\pi(\vc{x}) / q(\vc{x})$ is bounded and as nearly constant as possible, in much the same way that optimal proposal densities would be chosen for importance sampling or for rejection sampling. 
Liu \cite{liu:96} compares the efficiency of the three methods, concluding that the independence Metropolis sampling is asymptotically comparable to rejection sampling, but that independence Metropolis sampling is simpler to implement as it does not require knowledge of the envelope constant.

A number of methods have been developed for MH that use either local gradient information to make transitions to regions of high probability   \cite{geweke:89,martin:12} or to  allow better exploration of the target density than can be obtained by random-walk sampling   \cite{duane:87,cunha:98,dostert:06}. In several methods, the transition is based on an optimization step.   
The Multiple-Try method \cite{liu:00} uses local minimization along a line in state space to propose candidate states for updating one of the current population of states. Alternatively, Tjelmeland and  Hegstad \cite{tjelmeland:01} described a two-step transition in the Metropolis-Hastings method in which mode-jumping transitions alternated with mode-exploring transitions.

The Randomized Maximum Likelihood (RML) method \cite{oliver:96e} was  developed  as an independence Metropolis sampler for  the special case  in  which the target density $\pi(\vc{x}) $ is the product of a multivariate Gaussian prior $p(\vc{x})$ and a  likelihood function $p(\vc{d}|\vc{x})$ for which the errors in observations $\vc{d}$ are Gaussian.
The prior distribution for  $\vc{X}$ is assumed to be  multivariate normal  with mean $\boldsymbol{\mu}$ and covariance $\vc{C}_x$. 
Observations $\vc{d}\sbr{obs} = \vc{g}(\vc{x}) + \boldsymbol{\epsilon}_d$ with $\boldsymbol{\epsilon}_d \sim N(0,\vc{C}_d)$ are assimilated resulting in a posterior density 
\begin{equation} 
\pi(\vc{x}) \propto \exp \left( -\frac{1}{2} (\vc{x} - \boldsymbol{\mu} )\trp \vc{C}_x\inv (\vc{x} - \boldsymbol{\mu}) - \frac{1}{2} (\vc{g}(\vc{x})-\vc{d}\sbr{obs})\trp \vc{C}_d\inv (\vc{g}(\vc{x})-\vc{d}\sbr{obs}) \right)  .
\label{eq:pi_x}
\end{equation}  
Candidate samples in RML are obtained by a two-step process. In the first step, randomized samples from the prior distribution for model variables and data variables are generated. In the second step, the candidate state is obtained by minimization of a stochastic objective function.  
The distribution of candidate states $q(\vc{x})$ depends on both the model and data variables so computation of the transition probability requires marginalization,  $q(\vc{x}) = \int q(\vc{x},\vc{d}) \, d\vc{d}$.  For problems in which the prior is Gaussian, the observation operator is linear, and errors in observations are additive and Gaussian, the proposal density $q(\vc{x})$ is equal to the target density $\pi(\vc{x})$ so that all independently proposed candidate states  are accepted in Metropolis-Hastings. For problems in which the observation operator is nonlinear, proposed candidates are always placed in regions of high probability density, but evaluation of the Metropolis-Hastings acceptance test is difficult as it requires evaluation of the marginal density for model variables. Consequently, in practice the MH test is ignored for RML in large geoscience inverse problems \cite{emerick:13,gao:06a}. Bardsley et al.
\cite{bardsley:14} showed that for non-linear problems  in which the Jacobian of the observation operator is differentiable and monotonic, the MH acceptance test could be based on a Jacobian evaluated at the maximizer of Eq.~\ref{eq:pi_x}. In this case, the probability of candidate states could be relatively easily computed and sampling was shown to be correct for several test problems.

Here we describe a modified Metropolization approach in which the randomized data are included in the state vector.  The proposal of model variables is accomplished via a local optimization, but  calibrated data variables are retained in the state space and used for evaluation of the probability of acceptance of the state. By doing this, the method allows sampling from multimodal distributions and the need to compute the marginal density for model variables is avoided. Correct sampling requires  determination of the distribution of the proposals which is easier when the state space is augmented with data variables.

The paper is organized as follows. In Section~\ref{sec:old_rml} we summarize the original RML algorithm as an independence Metropolis sampler for which only model variables are included in the state of the Markov chain and discuss a pseudo-marginal MCMC approach to the sampling problem. In Section~\ref{sec:new_rml} we describe an augmented state version of the RML independence Metropolis sampler. For this version, computation of the Metropolis-Hastings ratio is simpler as it only requires the Jacobian of the transformation of the augmented state variables. Section~\ref{sec:examples} provides several examples for validation of the algorithm.

\section{RML for model variables} \label{sec:old_rml}

In both the quasi-linear estimation method  \cite{kitanidis:95} and the randomized maximum likelihood method \cite{oliver:96e}, samples are generated in high probability regions of the posterior pdf by the simple expedient of simultaneously minimizing the magnitude of misfit of the model with perturbed observations and the  magnitude of the change in the model variables from an unconditional sample. Unconditional realizations of models variables ($\vc{x}\sbr{uc}$) are sampled from the prior and realizations of observation ($\vc{d}\sbr{uc}$) are sampled from the distribution of model noise,  
\[ \vc{x}\sbr{uc} \sim N[ \boldsymbol{\mu}, \vc{C}_x] \quad  \text{ and } \quad
 \vc{d}\sbr{uc} \sim N[ \vc{d}\sbr{obs}, \vc{C}_d] . \]
A sample from a high probability region is generated by minimizing a nonlinear least-squares cost  function:
\begin{equation}%
\vc{x}_\ast = \argmin_{\vc{x}}   \Bigl[  \frac{1}{2} (\vc{x} -  \vc{x}\sbr{uc})\trp 
\vc{C}_x\inv (\vc{x} -  \vc{x}\sbr{uc})   
  +\frac{1}{2}  (\vc{g}(\vc{x}) - \vc{d}\sbr{uc})\trp \vc{C}_d\inv (\vc{g}(\vc{x}) - \vc{d}\sbr{uc})  \Bigr].
   \label{eq:fYus}
\end{equation}

Although the samples generated in this way tend to be located in parts of parameter space with high posterior probability, the samples are not necessarily distributed according to the posterior distribution. The distribution of RML samples can, however, be computed if the inverse transformation from the calibrated samples to the prior samples is known. 
If the objective function (Eq.~\ref{eq:fYus}) is differentiable, then a necessary (but not sufficient) condition for $\vc{x}_\ast$ to be a minimizer is that
\begin{equation} \vc{x}\sbr{uc} = 
 \vc{x}_\ast + \vc{C}_x \vc{G}\trp
\vc{C}_d\inv ( \vc{g}(\vc{x}_\ast ) - \vc{d}\sbr{uc} ) 
   \label{eq:Xus}
\end{equation}
where $\vc{G}\trp = \nabla_x \vc{g}$. 
Eq.~\ref{eq:Xus} provides the inverse transformation for the unconditional samples, restricted to the region of $(\vc{x}_\ast, \vc{d}\sbr{uc})$ for which $\vc{x}_\ast$ is a minimizer of Eq.~\ref{eq:fYus}.

The joint distribution for $\vc{x}\sbr{uc}$ and $\vc{d}\sbr{uc}$ is
\begin{equation*}%
f(\vc{x}\sbr{uc},\vc{d}\sbr{uc})  \propto  \exp \Bigl(  -\frac{1}{2} (\vc{x}\sbr{uc} - \boldsymbol{\mu})\trp 
\vc{C}_x\inv (\vc{x}\sbr{uc} - \boldsymbol{\mu})  
 -\frac{1}{2}  (\vc{d}\sbr{uc} - \vc{d}\sbr{obs})\trp \vc{C}_d\inv (\vc{d}\sbr{uc} - \vc{d}\sbr{obs})  \Bigr)
\end{equation*}
and hence the joint 
probability of proposing $(\vc{x}_\ast,\vc{d}\sbr{uc})$ is
\begin{equation}
	q(\vc{x}_\ast,\vc{d}\sbr{uc}) = f (\vc{x}\sbr{uc} (\vc{x}_\ast,\vc{d}\sbr{uc} ),\vc{d}\sbr{uc}) |\vc{J}|
	\label{eq:q_RML_old}
\end{equation}
where $\vc{J}$ is the Jacobian of the inverse transformation from $(\vc{x}_\ast,\vc{d}_\ast)$ to $(\vc{x}\sbr{uc},\vc{d}\sbr{uc})$. By choosing $\vc{d}_\ast = \vc{d}\sbr{uc}$, the Jacobian determinant requires only computation of 
\[
	|\vc{J}| = \biggl| \frac{\partial (\vc{x}\sbr{uc},\vc{d}\sbr{uc})}{\partial (\vc{x}_\ast)} \biggr|.
\]
The marginal probability of proposing $\vc{x}_\ast$ is then obtained by integrating 
$q(\vc{x}_\ast,\vc{d}\sbr{uc})$ over the data space,
\begin{equation}
	q(\vc{x}_\ast) = \int_{D} f (\vc{x}\sbr{uc} (\vc{x}_\ast,\vc{d}\sbr{uc} ),\vc{d}\sbr{uc}) \, |\vc{J}(\vc{x}_\ast,\vc{d}\sbr{uc} )| \, d\vc{d}\sbr{uc} .
\label{eq:marginal_old_RML}
\end{equation}

For an independence sampler in the Metropolis-Hastings algorithm \cite{tierney:94},  the probability of proposing a transition to state $\vc{x}_\ast$   is 
independent of the current state 
$\vc{x}$, then the proposed state  $\vc{x}_\ast$ is accepted with probability
\begin{equation}%
\alpha(\vc{x}, \vc{x}_\ast) = \min  \left( 1,
         \frac{\pi(\vc{x}_\ast ) q(\vc{x}) }{\pi(\vc{x}) q(\vc{x}_\ast) }   \right).
   \label{eq:hastingsaij}
\end{equation}
The 
probability density for the proposed model, $\pi(\vc{x}_\ast)$, is
computed from Eq.~\ref{eq:pi_x}.
Note that the probability $\pi$ is not based on the quality of the match to the perturbed data
obtained in the minimization, but on the quality of the match to the
prior model and the actual observed data.
Algorithm~\ref{alg:RML_old} describes the method for using RML as a proposal mechanism in a Metropolis-Hastings procedure \cite{oliver:96e}.

\begin{algorithm}
\caption{RML for model variables}
\label{alg:RML_old}
\begin{algorithmic}[1]
\State Generate initial state  $\vc{x}_0 \sim N[\boldsymbol{\mu} , \vc{C}_x]$ 
\For{$i \le i\sbr{max}$ }
\Procedure{Generate candidate state}{}
\State Generate $\vc{x}_{uc} \sim N[\boldsymbol{\mu} , \vc{C}_x]$ and  $\vc{d}_{uc} \sim N[\vc{d}\sbr{obs} , \vc{C}_d]$
  
\State Compute 
\State  $\vc{x}_\ast = \argmin_{\vc{x}}   \Bigl[  \frac{1}{2} (\vc{x} -  \vc{x}\sbr{uc})\trp 
\vc{C}_x\inv (\vc{x} -  \vc{x}\sbr{uc})   
  +\frac{1}{2}  (\vc{g}(\vc{x}) - \vc{d}\sbr{uc})\trp \vc{C}_d\inv (\vc{g}(\vc{x}) - \vc{d}\sbr{uc})  \Bigr]$

\EndProcedure
\State Compute  proposal density $q(\vc{x}_\ast)= \int_{D} q(\vc{x}_\ast,\vc{d}\sbr{uc}) \, d\vc{d}\sbr{uc} $  using Eq.~\ref{eq:q_RML_old} for $q(\vc{x}_\ast,\vc{d}\sbr{uc})$.
\State Compute   $\alpha(\vc{x}_{i},\vc{x}_\ast) = \min \left(1, \frac{\pi(\vc{x}_\ast) q(\vc{x}_{i}) }{\pi(\vc{x}_i) q(\vc{x}_\ast)} \right)$ 
\State Generate $u$ from $U(0,1)$
\If {$u \le \alpha(\vc{x}_{i},\vc{x}_\ast)$}
    \State $\vc{x}_{i+1} \gets \vc{x}_\ast$
\Else
    \State $\vc{x}_{i+1} \gets \vc{x}_i$
\EndIf
\EndFor
\end{algorithmic}
\end{algorithm}

When the relationship of model variables to observations is linear and observation errors are additive and Gaussian, it is straightforward to show that the minimizers of Eq.~\ref{eq:fYus} are distributed as the posterior distribution for $\vc{x}$ \cite{bardsley:13,oliver:96b} and hence
all  proposed transitions are accepted when Algorithm~\ref{alg:RML_old} is used  for Gauss-linear inverse problems.  

When Algorithm~\ref{alg:RML_old} was applied  to a problem for which the  posterior distribution was bimodal (as in Example 1, below), 
the acceptance rate for proposed independent transitions was still very high (76\%)  \cite{oliver:96e}. 
Although the method was shown to sample correctly, the work required for 
computing the marginal distribution for the candidate states made application of the full method impractical in real problems. Consequently, when the method has been applied to large problems, the MH acceptance test was omitted \cite{calverley:11,chen:14,gao:06a}, with the understanding that the sampling method is then only approximately correct.
 
The need to numerically compute the marginal density $q(\vc{x}_\ast)$ from $q(\vc{x}_\ast,\vc{d}\sbr{uc})$ in Algorithm 1 could potentially be avoided through application of the pseudo-marginal MCMC method  \cite{andrieu:09,sherlock:15}, which only requires an unbiased estimate of the marginal distribution. When the pseudo-marginal MCMC method as described in Algorithm 9.7 of \cite{reich:15} was applied  to Example~1 (Section~\ref{sec:bimodal})  using a using a single unbiased sample at each step, however, sampling  from a Markov chain of length 10\spr{5} was very poor.
Results would almost certainly be improved  by increasing the sample size used to represent the marginal distribution  \cite{sherlock:15}, but  because of the dependence of $x_\ast$ on $d\sbr{uc}$ in the RML method, it is not feasible to draw multiple $d\sbr{uc}$ for a given $x_\ast$. 
 
\section{Augmented state RML} \label{sec:new_rml}

Because the major challenge with the use of Metropolized RML as  in Algorithm~\ref{alg:RML_old} is the computation of the marginal probability of proposing the candidate model variables, we introduce a modification that  avoids the need for computation of the marginal proposal density by augmenting
 the state $\vc{x}_\ast$ with suitably defined data variables $\vc{d}_\ast$. Although auxiliary variables can improve sampling in several ways \cite{storvik:11},  the purpose of including auxiliary variables in this case is primarily to simplify computation of the MH acceptance criterion.

 \subsection{Posterior (target) distribution}

Consider the joint space of model variables and data in the case for which  the observation errors and the modelization errors can both be modeled as Gaussian. The prior distribution of model variables is assumed Gaussian with mean $\boldsymbol{\mu}$ and covariance $\vc{C}_x$; the prior distribution of observations is assumed to be Gaussian with mean $\vc{d}\sbr{obs}$ and covariance $(1-\gamma) \vc{C}_d$; modelization errors are Gaussian with mean 0 and covariance $ \gamma \vc{C}_d$.
 The posterior  joint probability of $(\vc{x}, \vc{d})$, obtained by combining states of information is 
\begin{multline}%
\pi(\vc{x},\vc{d})  \propto  \exp \Bigl[  -\frac{1}{2} (\vc{x} - \boldsymbol{\mu})\trp 
\vc{C}_x\inv (\vc{x} - \boldsymbol{\mu})  -\frac{1}{2 \gamma}  ( \vc{g}(\vc{x}) - \vc{d})\trp \vc{C}_d\inv (\vc{g}(\vc{x}) - \vc{d}) \\
  -\frac{1}{2 (1-\gamma)}  (\vc{d} - \vc{d}\sbr{obs})\trp \vc{C}_d\inv (\vc{d} - \vc{d}\sbr{obs})  \Bigr].
   \label{eq:post_md}
\end{multline}
After some rearrangement of terms, the joint probability can be factored into the product of the marginal posterior density for model variable $\vc{x}$ and the posterior density for $\vc{d}$ conditional to $\vc{x}$, 
\begin{multline}%
\pi(\vc{x},\vc{d})  \propto  \exp \Bigl[  -\frac{1}{2} (\vc{x} - \boldsymbol{\mu})\trp 
\vc{C}_x\inv (\vc{x} - \boldsymbol{\mu})  -\frac{1}{2}  ( \vc{g}(\vc{x}) - \vc{d}\sbr{obs})\trp \vc{C}_d\inv (\vc{g}(\vc{x}) - \vc{d}\sbr{obs}) \Bigr] \\
 \exp \Bigl[ -\frac{1}{2 \gamma (1-\gamma)}  \left(\vc{d} - \vc{g}(\vc{x}) + \gamma (\vc{g}(\vc{x}) - \vc{d}\sbr{obs}) \right)\trp      \vc{C}_d\inv  \left(\vc{d} - \vc{g}(\vc{x}) + \gamma (\vc{g}(\vc{x}) - \vc{d}\sbr{obs}) \right)  \Bigr].
   \label{eq:post_md_factored}
\end{multline} 
Although the introduction of Gaussian modelization error does not affect the posterior marginal distribution of $\vc{x}$ as long as it is compensated for by a reduction in observation error, the posterior  joint probability of $(\vc{x}, \vc{d})$ does depend strongly on the value of $\gamma$ as discussed in Section~\ref{sec:select_rho_gamma} on the selection of $\gamma$.

\subsection{Proposal distribution}

As in the original version of RML,
we draw  samples from the normal distributions
 $\vc{x}\sbr{uc} \sim N[ \boldsymbol{\mu} , \vc{C}_x]$ and
 $\vc{d}\sbr{uc} \sim N[ \vc{d}\sbr{obs} , \vc{C}_d]$ and denote the joint distribution as  $p(\cdot,\cdot)$.
Candidate transitions in a MH algorithm are obtained by minimizing a nonlinear least squares cost function
\begin{multline}%
(\vc{x}_\ast, \vc{d}_\ast) = \argmin_{\vc{x},\vc{d}}   \Bigl[  \frac{1}{2} (\vc{x} -  \vc{x}\sbr{uc})\trp 
\vc{C}_x\inv (\vc{x} -  \vc{x}\sbr{uc})     +\frac{1}{2 \rho}  (\vc{g}(\vc{x}) - \vc{d})\trp \vc{C}_d\inv (\vc{g}(\vc{x}) - \vc{d})  \\
  +\frac{1}{2 (1-\rho)}  (\vc{d} - \vc{d}\sbr{uc})\trp \vc{C}_d\inv (\vc{d} - \vc{d}\sbr{uc})  
 \Bigr].
   \label{eq:argmin_augmented}
\end{multline}
Implicit expressions for the relationship between $(\vc{x}_\ast, \vc{d}_\ast)$ and $(\vc{x}\sbr{uc},\vc{d}\sbr{uc})$ are derived from requirement that the gradient at the minimum must vanish. At the minimum, the following relationships hold:
\begin{equation} 
\vc{x}_\ast -  \vc{x}\sbr{uc}    +\frac{1}{\rho}  \vc{C}_x \vc{G}\trp \vc{C}_d\inv (\vc{g}(\vc{x}_\ast) - \vc{d}_\ast)  = 0  
  \label{eq:m}
 \end{equation}
and
\begin{equation}    \vc{d}_\ast  - \rho \, \vc{d}\sbr{uc}  
  -  (1- \rho)  \vc{g}(\vc{x}_\ast)   = 0.
  \label{eq:d}
\end{equation}
 Using Eq.~\ref{eq:d} to eliminate $ \vc{d}_\ast$ from Eq.~\ref{eq:m} gives
\begin{equation} 
 \vc{x}_\ast -  \vc{x}\sbr{uc}    +  \vc{C}_x \vc{G}\trp \vc{C}_d\inv (  \vc{g}(\vc{x}_\ast) -  \vc{d}\sbr{uc}  
      )  = 0  
  \label{eq:m2}
 \end{equation}
which shows that the marginal distribution of $\vc{x}_\ast$ is independent of $\rho$.
For Gauss-linear problems, $\vc{x}_\ast$ is distributed according to the posterior marginal distribution of $\vc{x}$. 

The inverse mapping,  obtained from Eqs.~\ref{eq:m} and~\ref{eq:d},
\begin{equation} 
 \begin{bmatrix} \vc{x}\sbr{uc} \\  \vc{d}\sbr{uc} \end{bmatrix} =  
 \begin{bmatrix} \vc{x}_\ast + \frac{1}{\rho} \vc{C}_x \vc{G}\trp \vc{C}_d\inv (\vc{g}(\vc{x}_\ast) - \vc{d}_
\ast)    \\
\frac{1}{\rho}  \, \vc{d}_\ast    
  -  \left(\frac{1- \rho}{\rho} \right)  \vc{g}(\vc{x}_\ast) 
  \end{bmatrix}.
  \label{eq:muc_duc}
 \end{equation}
is invertible if the domain of the mapping is restricted to the range of the mapping given by Eq.~\ref{eq:argmin_augmented} which we denote by $\Gamma$.
The distribution of candidate states, is then 
\begin{equation} 
q(\vc{x}_\ast,\vc{d}_\ast) = \begin{cases} p \left(  \vc{x}\sbr{uc} (\vc{x}_\ast, \vc{d}_\ast), \vc{d}\sbr{uc} (\vc{x}_\ast, \vc{d}_\ast) \right) |\vc{J}(\vc{x}_\ast, \vc{d}_\ast)|   & \text{if $(\vc{x}_\ast, \vc{d}_\ast) \in \Gamma$} \\
0  & \text{if $(\vc{x}_\ast, \vc{d}_\ast) \not \in \Gamma$}
\end{cases}
\label{eq:q_s_joint}
\end{equation}
where $\vc{x}\sbr{uc} (\vc{x}_\ast, \vc{d}_\ast)$ and $\vc{d}\sbr{uc} (\vc{x}_\ast, \vc{d}_\ast)$ are defined in Eq.~\ref{eq:muc_duc} and the dependence of the expressions on $\rho$ has been suppressed for clarity.

Because the state is composed of both model and data variables, the Jacobian matrix takes a block form with
\[ \begin{split}
\frac{\partial \vc{x}\sbr{uc}^\alpha}{\partial \vc{x}_\ast^\beta} & = I^{\alpha \beta} + \frac{1}{\rho} \sum_\gamma \sum_i \sum_j  \vc{C}_x^{\alpha \gamma}   \left[  \vc{G}^{i \gamma} \left[ \vc{C}_d\inv \right]^{ij} \vc{G}^{j \beta}   + \frac{\partial \vc{G}^{i \gamma}}{\partial \vc{x}^\beta}   \left[ \vc{C}_d\inv \right]^{ij}  (\vc{g}^j(\vc{x}_\ast) - \vc{d}^j_\ast)  ]  \right]
\end{split}
\]
and  
\[ \vc{G}^{i \alpha} = \frac{\partial \vc{g}^i}{\partial \vc{x}^\alpha} . \]
Other derivatives required for the Jacobian can be compactly written in matrix notation.
\[ \frac{\partial  \vc{x}\sbr{uc} }{ \partial  \vc{d}_\ast } =  - \frac{1}{\rho} \vc{C}_x \vc{G}\trp \vc{C}_d\inv ,
\]
\[ \frac{\partial  \vc{d}\sbr{uc} }{ \partial  \vc{x}_\ast } =  - \left(\frac{1- \rho}{\rho} \right)  \vc{G}
\]
and
\[ \frac{\partial  \vc{d}\sbr{uc} }{ \partial  \vc{d}_\ast } =   \frac{1}{\rho} I.
\]

For a linear Gaussian inverse problem, the relationship between the proposed states and the unconditional states is
\[
\begin{split}
\begin{bmatrix} \vc{x}_\ast \\ \vc{d}_\ast \end{bmatrix} 
& =
\begin{bmatrix}
\vc{C}_d \left( \vc{G} \vc{C}_x \vc{G}\trp + \vc{C}_d \right)\inv & 
\vc{C}_x \vc{G}\trp  \left( \vc{G} \vc{C}_x \vc{G}\trp + \vc{C}_d \right)\inv \\
(1-\rho )  \vc{G}\trp \vc{C}_d \left( \vc{G} \vc{C}_x \vc{G}\trp + \vc{C}_d \right)\inv &
\left( \vc{G}  \vc{C}_x \vc{G}\trp  + \rho  \vc{C}_d \right)   \left( \vc{G} \vc{C}_x \vc{G}\trp + \vc{C}_d \right)\inv \\
 \end{bmatrix}
 \begin{bmatrix} \vc{x}\sbr{uc} \\ \vc{d}\sbr{uc} \end{bmatrix} 
 \\
 & =
\vc{A} \begin{bmatrix} \vc{x}\sbr{uc} \\ \vc{d}\sbr{uc} \end{bmatrix} .
 \end{split}
 \]
The  covariance of the proposed states  $(\vc{x}_\ast, \vc{d}_\ast)$ is  
\[
\begin{split}
\vc{C}\sbr{post}
& =
\vc{A} \begin{bmatrix} \vc{C}_x & 0 \\ 0 & \vc{C}_d  \end{bmatrix} \vc{A}\trp 
\\
& =
 \begin{bmatrix}
\vc{C}_{x'} & 
  \vc{C}_{x'} \vc{G}\trp   \\
  \vc{G} \vc{C}_{x'}   &
  \vc{G} \vc{C}_{x'} \vc{G}\trp  + \rho^2 \vc{C}_d \vc{A}_D\inv \vc{C}_d
 \end{bmatrix}
 \end{split}
 \]
 where  
$ \vc{C}_{x'} =  \left( \vc{C}_x\inv   +  \vc{G}\trp \vc{C}_d\inv \vc{G} \right)\inv$ and $\vc{A}_D =  \left( \vc{G} \vc{C}_x \vc{G}\trp + \vc{C}_d \right)$.
 Note that for $\rho = 0$, the covariance of proposed states  is the same as the posteriori covariance for the joint model-data space of Tarantola \cite{tarantola:87}, although the covariance is then rank deficient.  

The joint density given by Eq.~\ref{eq:q_s_joint} can be used as the proposal density in an independence MH sampler for sampling of the joint posterior distribution    (Eq.~\ref{eq:post_md}) with the objective of sampling from the posterior marginal distribution for model variables (Eq.~\ref{eq:pi_x}). Algorithm~\ref{alg:RML_new} describes the use of minimization to provide  candidate states. 

\begin{algorithm}
\caption{Augmented state RML for model-data variables}
\label{alg:RML_new}
\begin{algorithmic}[1]
\State Generate $\vc{x}_{0} \sim N[\boldsymbol{\mu} , \vc{C}_x]$ and  $\vc{d}_{0} \sim N[\vc{d}\sbr{obs} , \vc{C}_d]$
\For{$i \le i\sbr{max}$ }
\Procedure{Generate candidate state}{}
\State Generate $\vc{x}\sbr{uc} \sim N[\boldsymbol{\mu} , \vc{C}_x]$ and  $\vc{d}\sbr{uc} \sim N[\vc{d}\sbr{obs} , \vc{C}_d]$
  
\State Compute 
\State  $(\vc{x}_\ast, \vc{d}_\ast) = \argmin_{\vc{x},\vc{d}}   \Bigl[  \frac{1}{2} (\vc{x} -  \vc{x}\sbr{uc})\trp 
\vc{C}_x\inv (\vc{x} -  \vc{x}\sbr{uc})   \Bigr.$
\State $ \qquad \Bigl.  \mbox{} +\frac{1}{2 \rho}  (\vc{g}(\vc{x}) - \vc{d})\trp \vc{C}_d\inv (\vc{g}(\vc{x}) - \vc{d})  
    +\frac{1}{2 (1-\rho)}  (\vc{d} - \vc{d}\sbr{uc})\trp \vc{C}_d\inv (\vc{d} - \vc{d}\sbr{uc})  
 \Bigr]$
 
\EndProcedure
\State Compute  proposal density $q(\vc{x}_\ast,\vc{d}_\ast)$  using Eq.~\ref{eq:q_s_joint}.
\State Compute   $\alpha(\vc{x}_{i}, \vc{d}_i,\vc{x}_\ast,\vc{d}_\ast) = \min \left(1, \frac{\pi(\vc{x}_\ast, \vc{d}_\ast) q(\vc{x}_{i},\vc{d}_i) }{\pi(\vc{x}_i,\vc{d}_i) q(\vc{x}_\ast,\vc{d}_\ast)} \right)$ 
\State Generate $u$ from $U(0,1)$
\If {$u \le \alpha(\vc{x}_{i}, \vc{d}_i,\vc{x}_\ast,\vc{d}_\ast)$}
    \State $\vc{x}_{i+1} \gets \vc{x}_\ast$ and $\vc{d}_{i+1} \gets \vc{d}_\ast$
\Else
    \State $\vc{x}_{i+1} \gets \vc{x}_i$ and $\vc{d}_{i+1} \gets \vc{d}_i$
\EndIf
\EndFor
\end{algorithmic}
\end{algorithm}

\subsection{Selection of $\rho$ and $\gamma$} \label{sec:select_rho_gamma}

The efficiency of the independence Metropolis sampler depends on two factors. The first is requirement that the proposal distribution $q$ (Eq.~\ref{eq:q_s_joint}) be close to the 
target density $\pi$ (Eq.~\ref{eq:post_md}) so that the ratio $\pi(\vc{x}_\ast, \vc{d}_\ast)/q(\vc{x}_\ast,\vc{d}_\ast)$ is as close to constant as possible \cite{tierney:94}. The second requirement is loosely that the proposal density be positive wherever the target density is positive so that the Markov chain is irreducible.

It is convenient to compare the proposal densities and the target densities based on the distribution of $\vc{d}$ given $\vc{x}$. From Eq.~\ref{eq:post_md_factored}, the mode of the target density for $\vc{d}$ is located at 
\begin{equation}
 \vc{d} = \vc{g}(\vc{x}) - \gamma (\vc{g}(\vc{x}) - \vc{d}\sbr{obs}) 
 \label{eq:mode_target}
 \end{equation}
and the covariance of $\vc{d}$ given $\vc{x}$ in the target distribution is $\gamma (1-\gamma) \vc{C}_d$.
Similarly, it is possible to factor the proposal density Eq.~\ref{eq:q_s_joint} as the product of the  target density for $\vc{x}$, a Gaussian distribution for $\vc{d}_\ast$ given $\vc{x}_\ast$,  term that depends on  $\vc{d}_\ast$, and the Jacobian determinant,
\begin{equation}  \begin{split}   
q(\vc{x}_\ast,\vc{d}_\ast)  & \propto  \exp \left[-\frac{1}{2} \left(\vc{x}_\ast -\boldsymbol{\mu} \right)\trp  \vc{C}_x\inv \left(\vc{x}_\ast - \boldsymbol{\mu} \right) 
- \frac{1}{2}    \left(  \vc{g}(\vc{x}_\ast) -\vc{d}\sbr{obs} \right)\trp \vc{C}_d\inv \left( \vc{g}(\vc{x}_\ast) -\vc{d}\sbr{obs} \right) \right] \\
& \quad \times  \exp \left[- \frac{1}{2 \rho^2} \left( \vc{d}_\ast- \vc{g}(\vc{x}_\ast) +  \rho \vc{V}\inv \boldsymbol{\eta}(\vc{x}_\ast)  \right)\trp \vc{V} \left( \vc{d}_\ast- \vc{g}(\vc{x}_\ast) +  \rho \vc{V}\inv \boldsymbol{\eta}(\vc{x}_\ast)  \right) \right] \\  
& \qquad \qquad \times  \exp \left[ \frac{1}{2} \boldsymbol{\eta}(\vc{x}_\ast)\trp \vc{V}\inv \boldsymbol{\eta}(\vc{x}_\ast) \right] \, |\vc{J}(\vc{x}_\ast,\vc{d}_\ast) |
\end{split} 
\label{eq:simplify_qXD}
\end{equation}
where 
\begin{equation} \vc{V} =  \vc{C}_d\inv + \vc{C}_d\inv \vc{G} \vc{C}_x \vc{G}\trp  \vc{C}_d\inv \end{equation}
and 
\begin{equation}   
\boldsymbol{\eta}(\vc{x}_\ast)  = \vc{C}_d\inv \left[ (\vc{G}  \vc{x}_\ast - \vc{g}(\vc{x}_\ast) ) -( \vc{G}\boldsymbol{\mu}  - \vc{d}\sbr{obs})    \right]  .
\end{equation}
Note that when the observation operator $\vc{g}$ is linear, $\boldsymbol{\eta}$, $\vc{V}$, and $\vc{J}$ are all independent of $\vc{x}_\ast$ and $\vc{d}_\ast$.

\begin{figure}[htbp!]
    \centering
    \begin{subfigure}[t]{0.45\textwidth}
        \centering
        \includegraphics[width=0.95\textwidth]{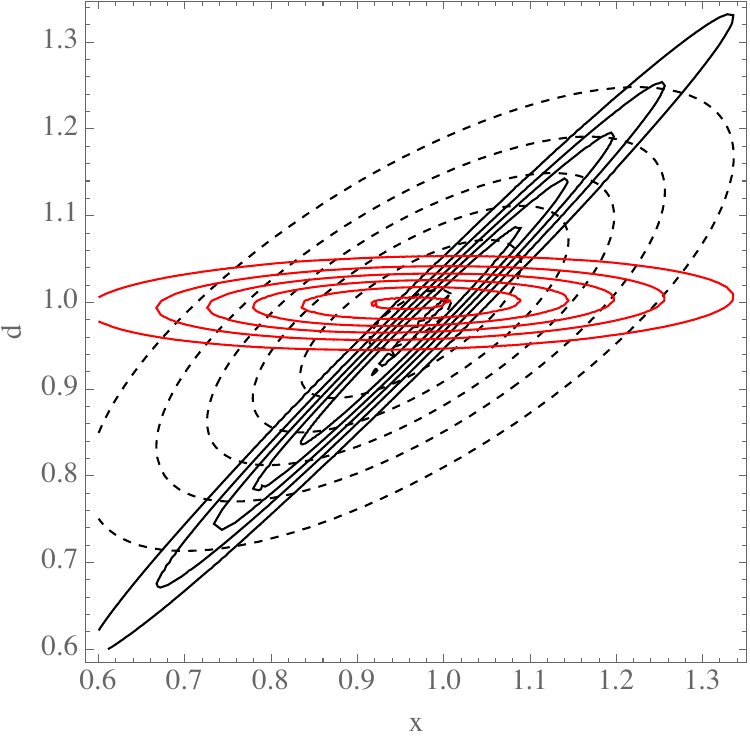} 
        \caption{Joint target pdf for three values of $\gamma$: 0.02 (black), 0.5 (dashed), 0.98 (red).} \label{fig:linear_compare_gamma_rho_a}
    \end{subfigure}%
    ~ 
    \begin{subfigure}[t]{0.45\textwidth}
        \centering
        \includegraphics[width=0.95\textwidth]{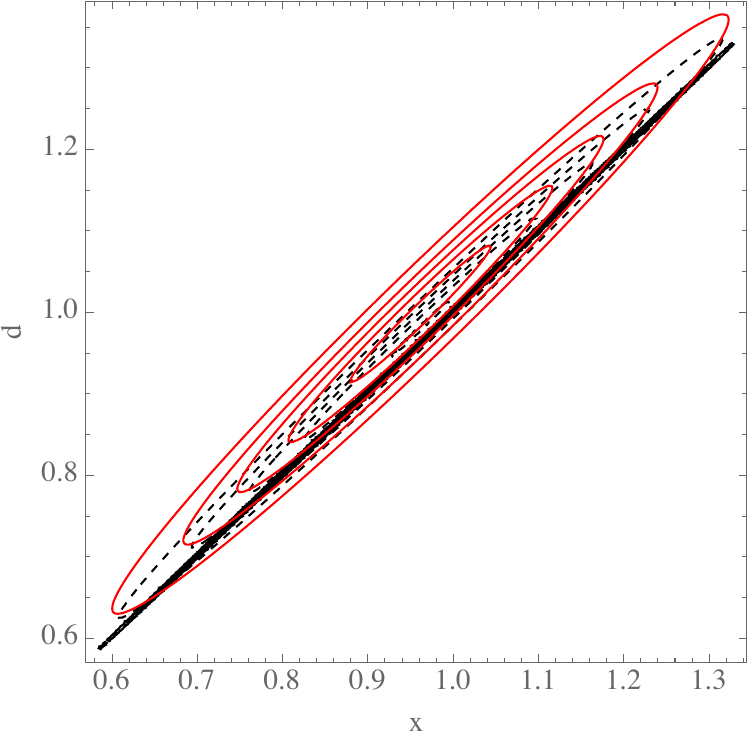} 
        \caption{Joint proposal pdf for three values of $\rho$: 0.05 (black), 0.5 (dashed), 0.95 (red).} \label{fig:linear_compare_gamma_rho_b}
    \end{subfigure}%
    \caption{Effect of  the values of $\rho$ and $\gamma$ on the joint target and proposal densities for a linear observation.}
    \label{fig:linear_compare_gamma_rho}
\end{figure}

For linear observations, both the target density and the proposal density are Gaussian in which case it is sufficient to compare the modes and the variances of the two distributions for selection of $\rho$ and $\gamma$.  The modes of the target distribution and the mode of the proposal distribution in this case  are equal when 
\[ -\rho \vc{V}\inv \boldsymbol{\eta} = \gamma (\vc{G} \vc{x}_\ast - \vc{d}\sbr{obs}) \]
and the variances of the two distributions are equal when 
\[ \rho^2 \vc{V}\inv  = \gamma (1-\gamma) \vc{C}_d . \]
For linear $g$, it is therefore not possible to satisfy both relationships exactly, except by choosing $\rho = \gamma = 0$, in which case the distributions are degenerate. 
Figure~\ref{fig:linear_compare_gamma_rho} compares the joint distributions for several values of $\gamma$ and $\rho$ in the case of a linear observation. It is clear that the two distributions will only be similar if $\gamma$ is small.

For nonlinear problems, the mapping (Eq.~\ref{eq:muc_duc}) from $(\vc{x}\sbr{uc} , \vc{d}\sbr{uc})$ to $(\vc{x}_\ast , \vc{d}_\ast)$  is not invertible unless the domain of the mapping is restricted.  In this case, it is necessary to select $\rho$ and $\gamma$ in such a way that the proposal density  is concentrated in the region where the mapping is bijective (and hence the proposal density is positive). Insight can be obtained by considering the case for which a single model variable and a single data variable. 

If $\partial^2 g/\partial x_\ast^2 \ne 0$, then solving for the location of the boundary, we obtain
\begin{equation}
 d_\ast = g(x_\ast) + C_d  \left( \frac{\partial^2 g}{\partial x_\ast^2} \right)\inv \left( \rho C_x\inv + G\trp C_d\inv G \right)   .
 \label{eq:boundary_proposal}
 \end{equation}
Increasing the  value of $\rho$ shifts the boundary away from $d_\ast = g(x_\ast)$ in the direction given by the sign of the second derivative of $g$.

The distance between the mode of the target density (Eq.~\ref{eq:mode_target}) and the boundary of the region of positive proposal density (Eq.~\ref{eq:boundary_proposal}) is 
\begin{equation*}
 \Delta d = C_d \left( \frac{d^2 g}{d x^2} \right) ( \rho C_x\inv + G\trp C_d\inv G) + \gamma (g(x) - d\sbr{obs}).
 \end{equation*}
The probability mass outside of the domain of validity is determined also by the variance of the distribution. The fraction of the cumulative distribution that is outside of the  domain of $q(\vc{x}_\ast,\vc{d}_\ast)$ is less than $\epsilon$ if 
\[ \Phi\left( \frac{\Delta d}{\sqrt{\gamma (1-\gamma) C_d }} \right) > 1 - \epsilon \, . \]
Thus for highly nonlinear problems, it is necessary to choose $\gamma \ll 1$ and $\rho \approx 1$ to satisfy the requirement that all states with significant probability density can be reached.

\begin{figure}[htbp!]
    \centering
    \begin{subfigure}[t]{0.31\textwidth}
        \centering
        \includegraphics[width=0.95\textwidth]{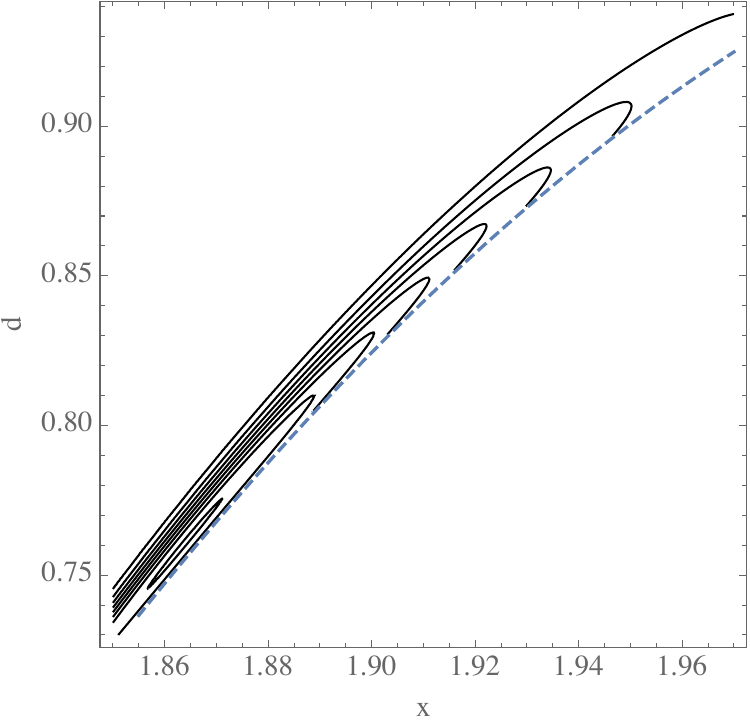} 
        \caption{One branch of $q(x,d)$ for $\rho=0.5$.} \label{fig:nonlinear_compare_gamma_rho_a}
    \end{subfigure}%
    ~ 
    \begin{subfigure}[t]{0.31\textwidth}
        \centering
        \includegraphics[width=0.95\textwidth]{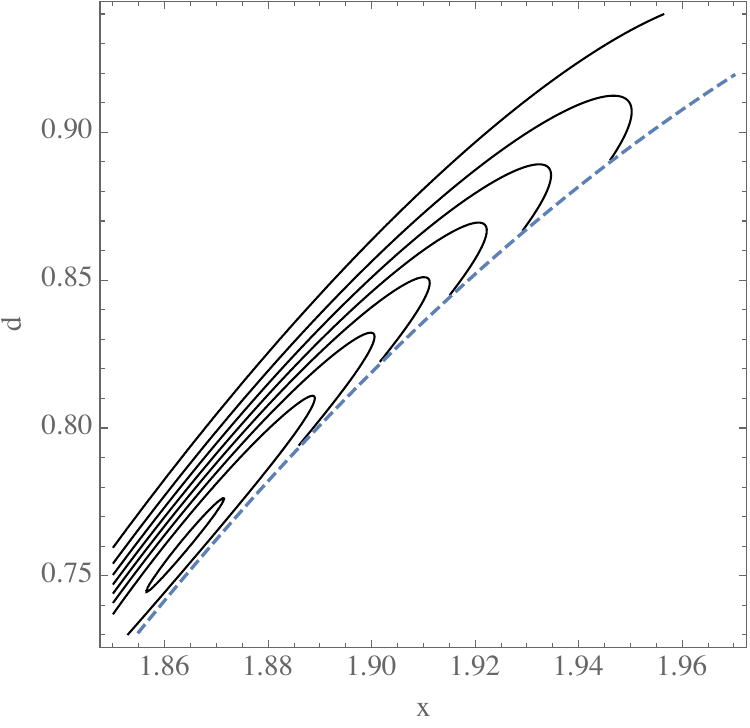} 
        \caption{One branch of $q(x,d)$ for $\rho=0.999$.} \label{fig:nonlinear_compare_gamma_rho_b}
    \end{subfigure}%
    ~
    \begin{subfigure}[t]{0.31\textwidth}
        \centering
        \includegraphics[width=0.95\textwidth]{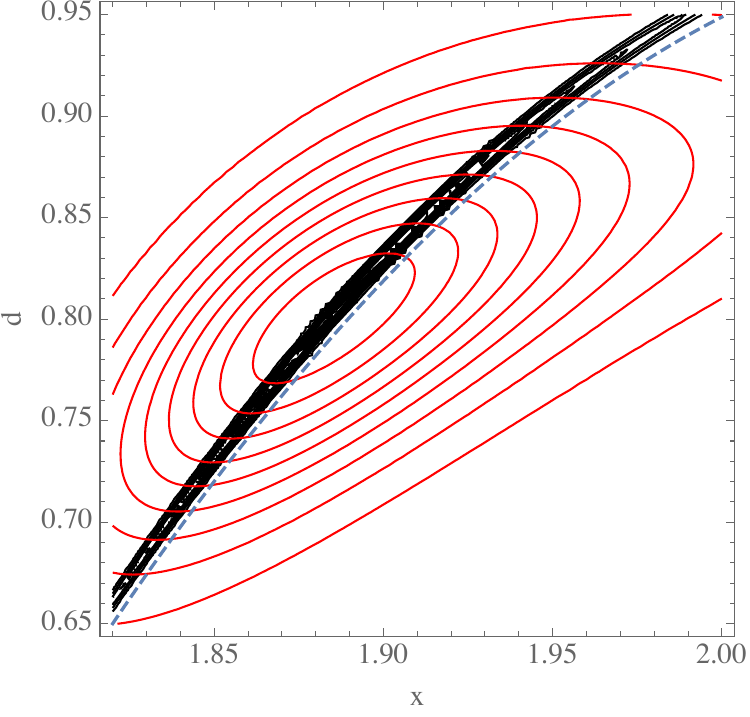} 
        \caption{One branch of $\pi(x,d)$ for two values of $\gamma$: 0.001 (black), 0.5 (red). Dashed blue contour shows boundary of proposal region.} \label{fig:nonlinear_compare_gamma_rho_c}
    \end{subfigure}%
    \caption{Effect of  the values of $\rho$ and $\gamma$ on the joint target and proposal densities for a quadratic observation.}
    \label{fig:nonlinear_compare_gamma_rho}
\end{figure}

Figure~\ref{fig:nonlinear_compare_gamma_rho} illustrates the effect of the choice of $\rho$ on the proposal density for the numerical example in Section~\ref{sec:bimodal}. Note that increasing $\rho$ from 0.5 (Fig.~\ref{fig:nonlinear_compare_gamma_rho_a}) to 0.999 (Fig.~\ref{fig:nonlinear_compare_gamma_rho_b}) increases the width of the proposal density but more importantly, it shifts the location of the boundary of the region with positive proposal density (dashed blue contour) away from the region of significant target density (Fig.~\ref{fig:nonlinear_compare_gamma_rho_c}). 

We note that although the RML method for generating proposals via the minimization of a cost function is somewhat inflexible, the target density for $(\vc{x},\vc{d})$ can be chosen much more freely as long as the marginal density for $\vc{x}$ is correct. In the terminology of Papaspiliopoulos et al. \cite{papaspiliopoulos:07}, we have chosen to use a centered parameterization, but if we had chosen $\tilde{\vc{d}} = \vc{d} -(1-\gamma) \vc{g}( \vc{x})$, then the pair $(\vc{x},\tilde{\vc{d}})$  would have been conditionally independent. That parameterization is similar to the choice made by \cite{oliver:14,wang:16}.  It does not alter the central difficulty, however, of ensuring that the target density is approximately zero when the proposal density is zero.  

\section{Examples} \label{sec:examples}

This section illustrates the  characteristics of the Metroplolized RML method on problems with multiple modes. In each case, it is possible to compare the distribution of samples with the true posterior probability density for a small inverse problem.  The first example consists of a single model variable and a single data variable so that the joint augmented distribution can be visualized easily. The posterior distribution on the model variable is bimodal but the probability in the region between modes is relatively large. In the second example,  the probability mass is located in more than 100 widely separated modes and the efficiency of the Metropolized RML method is compared with the preconditioned Crank Nicholson method  \cite{cotter:13a}. The third example illustrates application to a problem with nongaussian prior.

\subsection{Example 1: Bimodal} \label{sec:bimodal}

In this example, the posterior distribution for the model variable is bimodal. A region of small but significant probability density spans the region between the peaks.  This problem was originally used as to test the sampling of RML algorithm \cite{oliver:96e} where it was shown that correct sampling could be attained if the marginal distribution of transition proposals (Eq.~\ref{eq:marginal_old_RML}) is used in the MH test.
Computation of the marginal distribution was relatively difficult, however, even for this single-variable problem because the integration had to be limited to the region of the joint model-data space in which the mapping was invertible. 
Here we apply the augmented variable RML method (Algorithm \ref{alg:RML_new}) on the joint model-data space, without need for computation of the marginal distribution of proposals. The parameter $\gamma$ that determines the relative contribution of modelization error vs observation error in the posterior is set at a small nonzero value (0.01) to prevent degeneracy of the posterior distribution. 

\begin{figure}[htbp!]
    \centering
    \begin{subfigure}[t]{0.31\textwidth}
        \centering
        \includegraphics[width=0.95\textwidth]{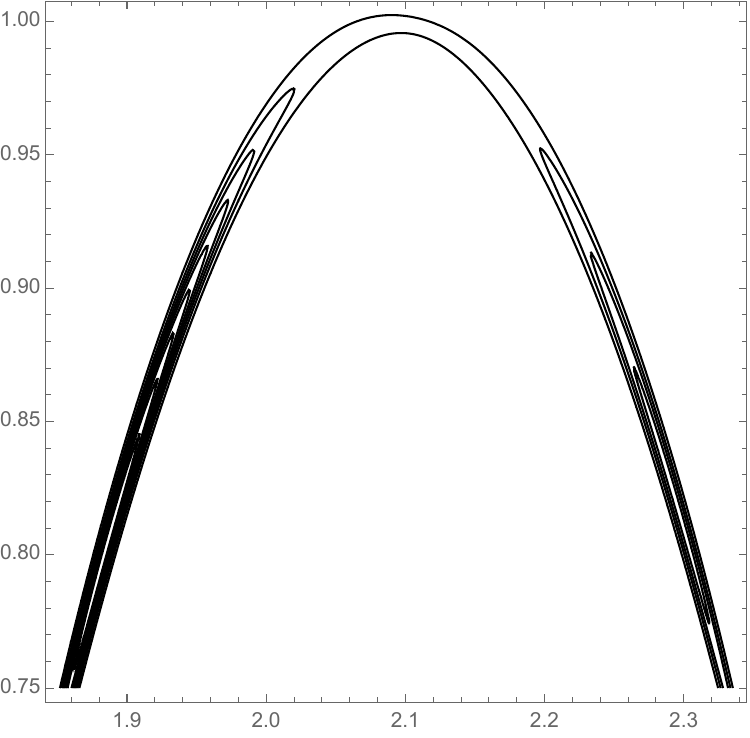} 
        \caption{Target joint density  ($\gamma=0.005$).} \label{fig:bimodal_compare_a}
    \end{subfigure}%
    ~ 
    \begin{subfigure}[t]{0.31\textwidth}
        \centering
        \includegraphics[width=0.95\textwidth]{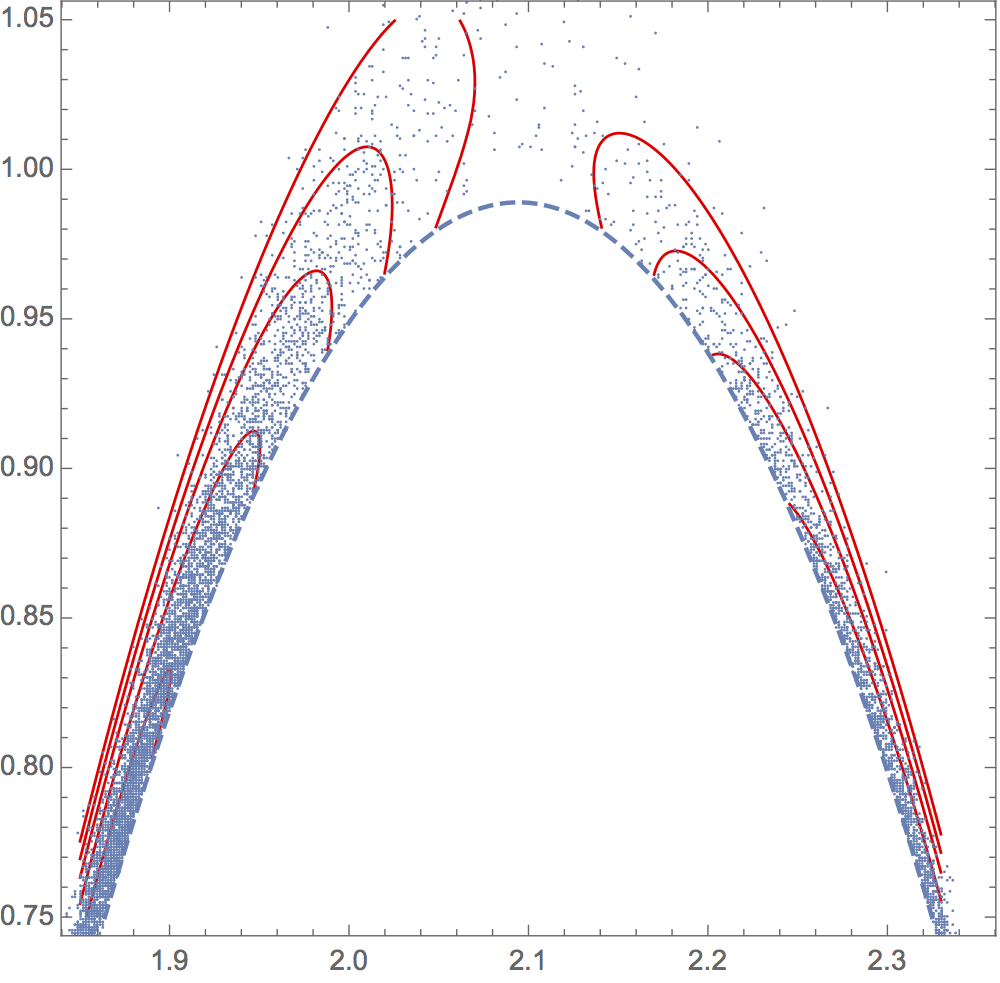} 
        \caption{RML proposal density ($\rho=0.995$).} \label{fig:bimodal_compare_b}
    \end{subfigure}%
   ~ 
    \begin{subfigure}[t]{0.31\textwidth}
        \centering
        \includegraphics[width=0.95\textwidth]{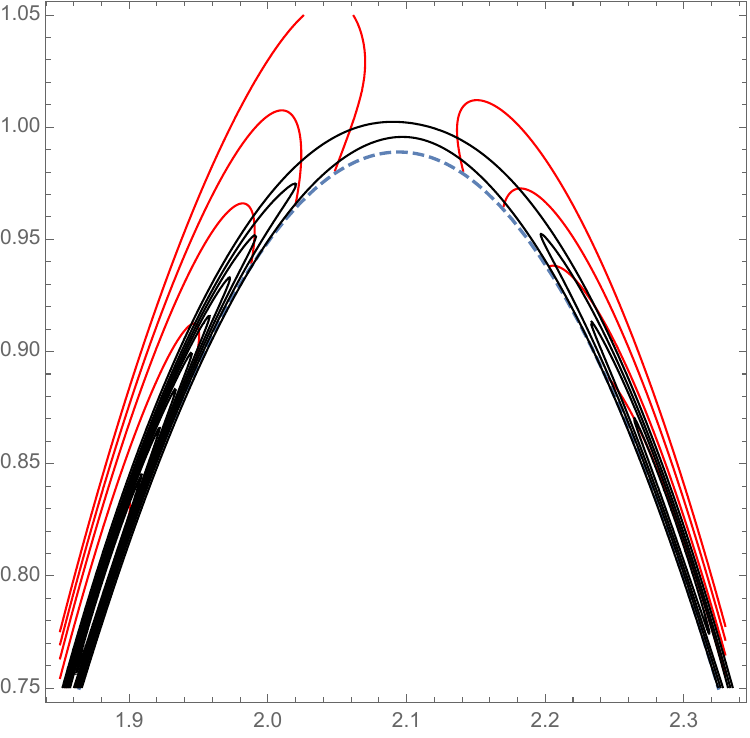} 
        \caption{Compare joint proposal to target.} \label{fig:bimodal_compare_c}
    \end{subfigure}%
    \caption{Comparison of RML proposal density with the target density for Example 1 (bimodal) sampling problem. Model variable (horizontal axis), data variable (vertical).}
    \label{fig:bimodal_compare_q_vs_pi}
\end{figure}

\begin{figure}[htbp!]
    \centering
    \begin{subfigure}[t]{0.45\textwidth}
        \centering
        \includegraphics[width=0.95\textwidth]{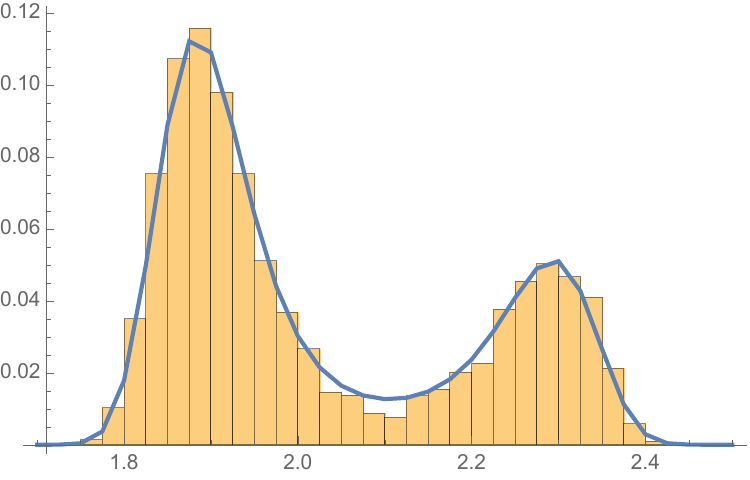} 
        \caption{Distribution of samples of model variable.} \label{fig:bimodal_results_a}
    \end{subfigure}%
    ~ 
    \begin{subfigure}[t]{0.45\textwidth}
        \centering
        \includegraphics[width=0.95\textwidth]{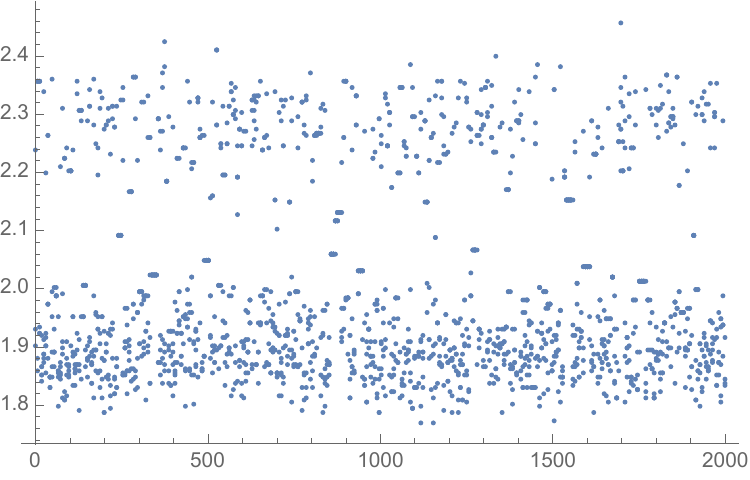} 
        \caption{Markov chain for model variable.} \label{fig:bimodal_results_b}
    \end{subfigure}%
    \caption{Metropolized RML sampling for Example 1 (bimodal pdf).}
    \label{fig:bimodal_results}
\end{figure}

We consider 
the  problem of  
sampling from the following univariate distribution 
\begin{equation}
	\pi (x) = a \exp \Bigl[ -\frac{(x- \mu)^{2}}{2 \sigma_{x}^{2}}
	    -\frac{(g(x)-d\sbr{obs})^{2}}{2 \sigma_{d}^{2}}\Bigr],
	\label{eq:pi}
\end{equation}
where $\mu=1.9$, $d\sbr{obs}=0.8$, $\sigma_{x}^{2}= 0.1$, 
$\sigma_{d}^{2}=0.01$, $g(x) = 1 - 9 (x- 2 \pi/3)^2 /2$, and $a \approx 4.567$. The true target distribution is shown in Fig.~\ref{fig:bimodal_results_a} as a solid curve.
 The first term in 
Eq.~\eqref{eq:pi} is the  prior 
density for m.  The second term  represents  the likelihood term in the  Bayesian inverse problem. Because of the nonlinearity of the observation operator $g(\cdot)$, the posterior is bimodal. 
An optimal  value of $\rho$ was determined by trial and error, but in fact, the MH acceptance rate is not highly sensitive to the value of $\rho$. In the interval $0.5 \le \rho \le 0.8$ the acceptance rate for RML proposals is between 62\% and 64 \%.  with a maximum acceptance rate of almost 0.64 when $\rho = 0.65$, which was the value we used.
For the choice  $\rho = 0.65$ and $\gamma = 0.01$, the joint proposal density for RML (Fig.~\ref{fig:bimodal_compare_a}) is similar to the posterior joint density of $x$ and $d$ (Fig.~\ref{fig:bimodal_compare_b}). Differences can more easily be seen in Fig.~\ref{fig:bimodal_compare_c} where the two pdfs are superposed. The main deficiency in the proposal density is a reduced probability of proposals in the region between the two peaks.

Figure \ref{fig:bimodal_results_a} shows the marginal posterior distribution of samples obtained using RML (Algorithm \ref{alg:RML_new}) compared with the target distribution (Eq.~\ref{eq:pi}) for a chain of length 40,000. Figure \ref{fig:bimodal_results_b} shows that the  mixing of the chain is  very good, as should be expected for an independence sampler with a high acceptance rate. 

\subsection{Example 2: Multimodal}

\begin{figure}[htbp!]
    \centering
    \begin{subfigure}[t]{0.45\textwidth}
        \centering
        \includegraphics[width=0.95\textwidth]{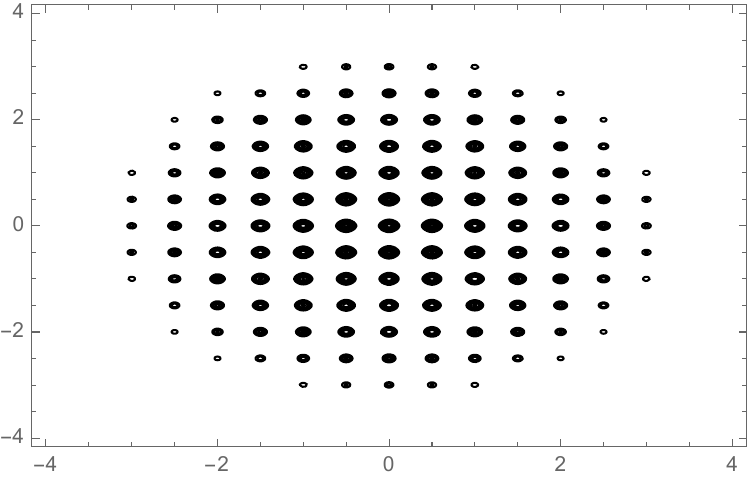} 
        \caption{Posterior pdf for model variables} \label{fig:posterior_many_a}
    \end{subfigure}%
    ~ 
    \begin{subfigure}[t]{0.45\textwidth}
        \centering
        \includegraphics[width=0.95\textwidth]{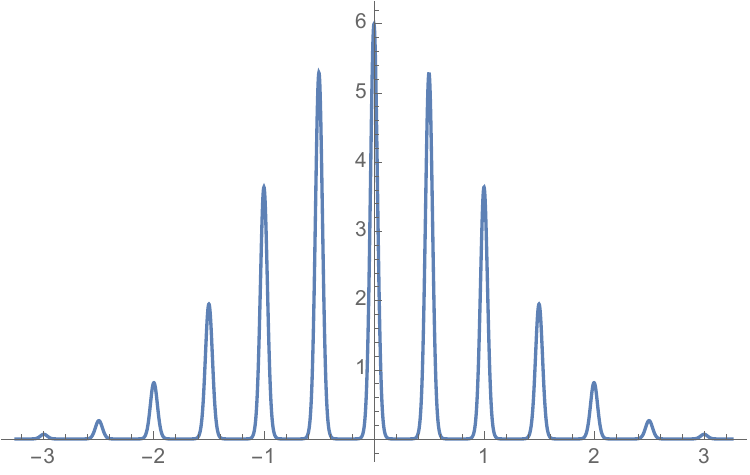} 
        \caption{Posterior pdf for model variables conditional to $x_2=0$} \label{fig:posterior_many_b}
    \end{subfigure}%
    \caption{True posterior pdf for a problem with many modes.}
    \label{fig:posterior_many}
\end{figure}

In this example, the prior distribution for the two model variables is independent standard normal. The observations are related to model variables as 
\[ \vc{g}(x_1, x_2) = \begin{bmatrix} \sin(2 \pi x_1) \\ 
  \sin( 2 \pi x_2)
  \end{bmatrix} .
  \]
The observations are  $\vc{d}\sbr{obs} = (0., 0.)$ with variance of observation error  equal 0.04.  The  posterior pdf for $\vc{x}$  is composed of many widely separated peaks making random walk methods relatively inefficient (Fig.~\ref{fig:posterior_many}).

\begin{figure}[htbp!]
    \centering
    \begin{subfigure}[t]{0.45\textwidth}
        \centering
        \includegraphics[width=0.95\textwidth]{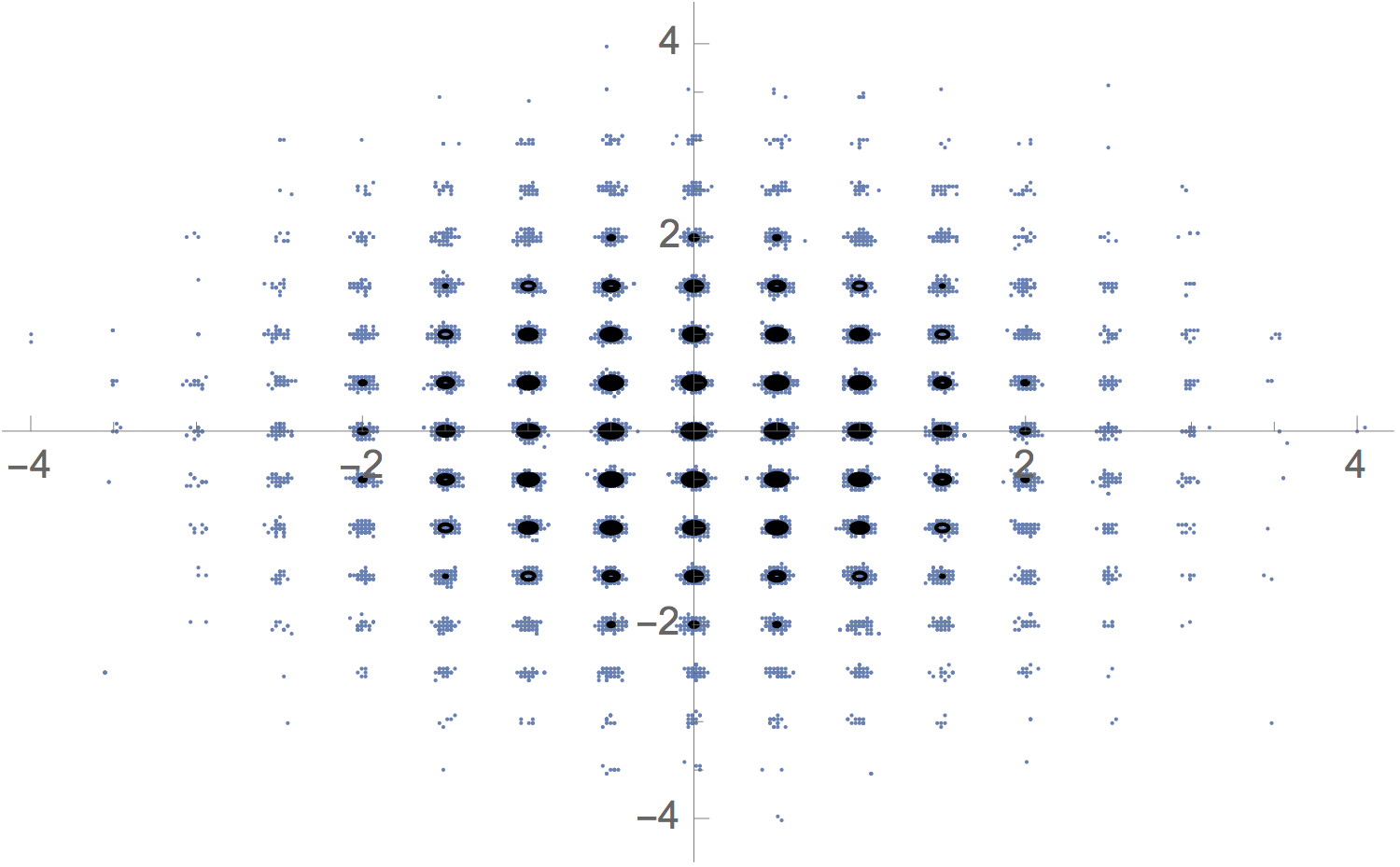} 
        \caption{All samples from MCMC length 40,000.} \label{fig:posterior_compare_a}
    \end{subfigure}%
    ~ 
    \begin{subfigure}[t]{0.45\textwidth}
        \centering
        \includegraphics[width=0.95\textwidth]{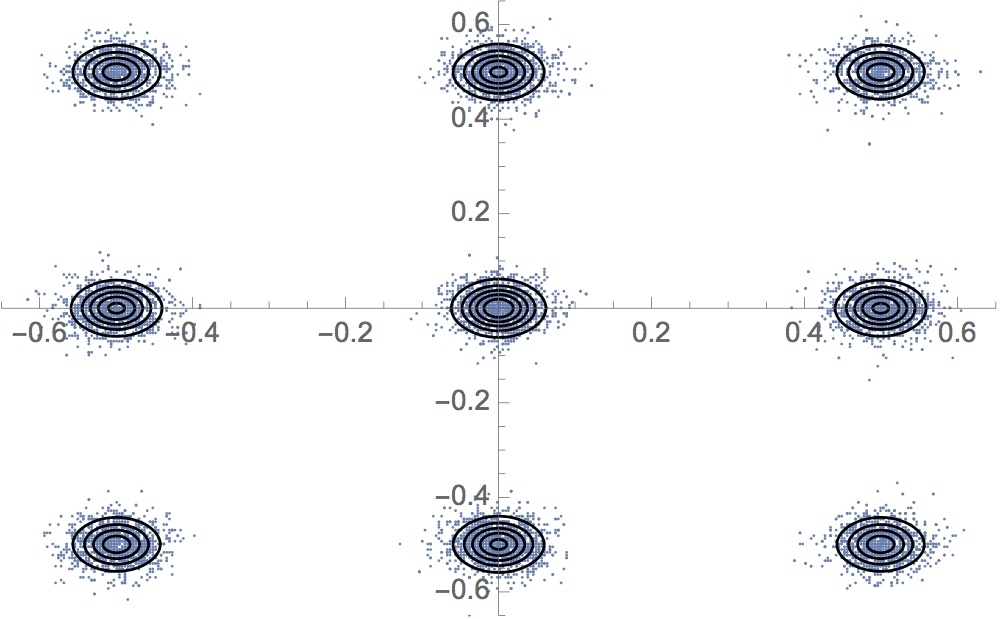} 
        \caption{Samples from MCMC in the neighborhood of the origin.} \label{fig:posterior_compare_b}
    \end{subfigure}%
    \caption{True posterior pdf for a problem with many modes.}
    \label{fig:posterior_compare}
\end{figure}

The Metropolized RML method (Algorithm~\ref{alg:RML_new}) was used for sampling with parameters $\rho=0.995$ and $\gamma = 0.005$. Figure~\ref{fig:posterior_compare_a} shows the  first 40000 elements of the Markov chain and Fig.~\ref{fig:posterior_compare_b} shows the distribution of samples from the same Markov chain in a more restricted region. Contours of the true pdf are plotted on top of the samples for comparison.
For this particular example, the acceptance rate was essentially independent of $\rho$ over a large range, varying between 87.1\% for $\rho=0.02$  to 87.4\% for $\rho=0.995$. 

\begin{figure}[htbp!]
\begin{center}
\begin{tabular}{cc}
    \raisebox{1ex}{\rotatebox{90}{Metropolized}}  & \includegraphics[width=0.95\textwidth]{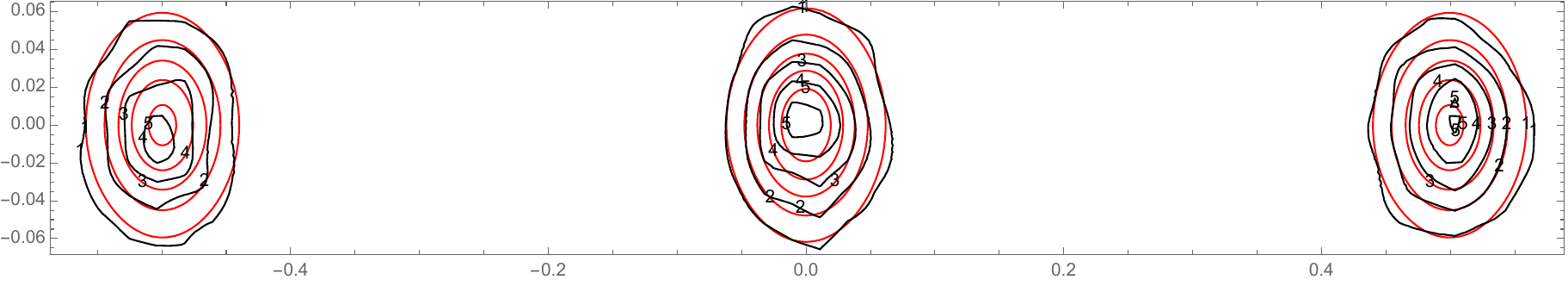} \\
  \raisebox{1ex}{\rotatebox{90}{Neglect MH}}   & \includegraphics[width=0.95\textwidth]{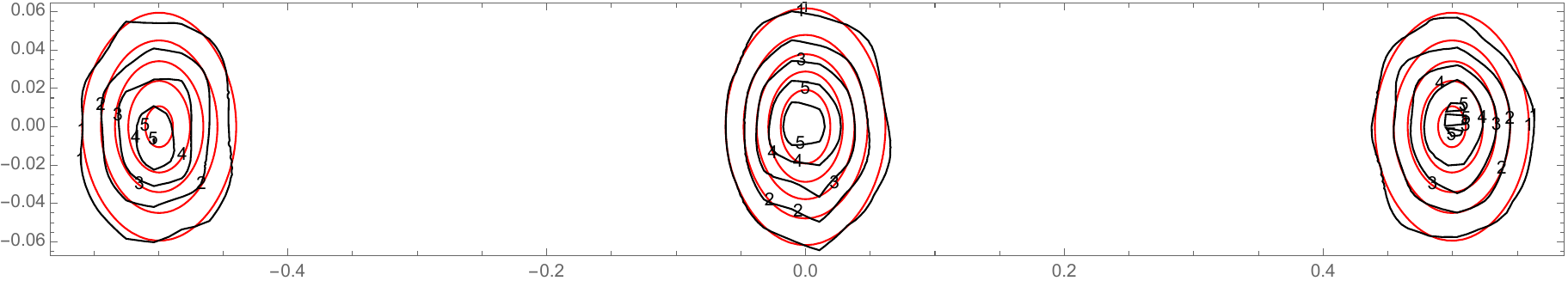}
\end{tabular}
\end{center}
    \caption{Compare estimated pdf (black contours) to the true pdf (red contours) for Metropolized RML (top) and RML without an acceptance test (bottom). Both Markov chains were of length 40,000 but only about 4000 samples are within this window.}
    \label{fig:compare_MRML_RML}
\end{figure}

Because the acceptance rate for independent proposals is so high in this example, it should be expected that  the proposal density for the model variables is quite similar to the target density. Fig.~\ref{fig:compare_MRML_RML} compares a section of the estimated pdf from the Metropolized RML to the true pdf (top) and a section of the estimated pdf from  RML without an acceptance test to the true pdf (bottom). The estimated pdf was obtained using a Gaussian smoothing kernel with bandwidth 0.01. The differences between the two sampled distributions are clearly quite subtle.  

\subsubsection{Computational cost}

Although the acceptance rate for independent samples in Metropolized RML method is quite high, the cost of generating proposals is substantially higher than the cost of a standard MCMC method. Efficient implementation generally requires estimation of gradients of the cost function and a reasonable choice for the initialization of the minimization.  The computational effort to generate a single proposal in RML for this numerical example  is approximately 15 function evaluations using a modified Levenberg-Marquardt minimization with finite difference estimation of gradients (MINPACK subroutine LMDIF) using $\vc{x}_{uc}$ to initialize the minimization.  In addition, computation of the Jacobian of the mapping requires an additional 5 function evaluations, hence the cost is approximately 20 function evaluations per element of the Markov chain. The acceptance rate for MH is 0.873 so the cost  is approximately $20/0.873 \approx 23$ functions evaluations per independent sample from the target distribution.

For comparison, we also applied the preconditioned Crank Nicholson method (pCN)  \cite{cotter:13a} to this example. The transition kernel for pCN is of the form 
\[ 
v^{(k)} = \sqrt{1-\beta^2}\, u^{(k)} + \beta \, \xi^{(k)},  \qquad \xi^{(k)} \sim N(0,C) 
\]
where $\beta$ is a tunable parameter of the algorithm and $u^{(k)}$ is the state of the Markov chain at step $k$. If one uses  $\beta = 0.055$ for an acceptance rate of 0.234 as recommended, a Markov chain of length $2 \times 10^6$  does not mix outside of a single mode. As an alternative, we selected an optimal value of  $\beta$ based on an estimate of effective sample size  using the algorithm in  Hoffman and Gelman  \cite{hoffman:14} with 5 independent Markov chains. For this problem with multiple widely separated modes, it was determined that  $\beta = 1$ was  optimal. Note that in this case pCN is an independence sampler. The acceptance rate for the pCN method was 0.022, which is equivalent to 46 function evaluations per independent sample, thus Metropolized RML is approximately twice as efficient.

If the standard deviation of noise in the data is reduced from 0.2 to 0.1, the modes of the posterior pdf become narrower and more widely separated. The acceptance rate for the pCN method falls to 0.0057 which is equivalent to 177 function evaluations per independent sample. 
The acceptance rate for Metropolized RML is slightly higher  (0.886) than when $\sigma_D = 0.2$  but the average number of function evaluations per iteration (15 + 5) is almost unchanged from that case. Thus, although the modes of the target pdf are more widely separated, the cost of  using Metropolized  RML is still approximately  $20/0.886 \approx 23$ function evaluations  per independent sample and the RML method becomes approximately 8.7 times as efficient as the pCN method.

For applications of RML to large-scale subsurface flow problems with many data, it has been reported  that although Newton-like methods typically converge quickly, the computational cost for estimation of the Hessian at each iteration makes Newton-like methods inefficient  compared to methods that only require computation of the gradient.  In a comparative study, Zhang and Reynolds \cite{zhang:02a} found the limited memory BFGS minimizer to be more efficient than other methods, including  Gauss-Newton, Modified Levenberg- Marquardt, preconditioned conjugate gradient, or BFGS. In all cases, the sample from the prior, $\vc{x}\sbr{uc}$, was used as the  initial guess for minimization.

\subsection{Example 3: Nongaussian prior}

\begin{figure}[htbp!]
    \centering
\begin{overpic}[width=0.45\textwidth]{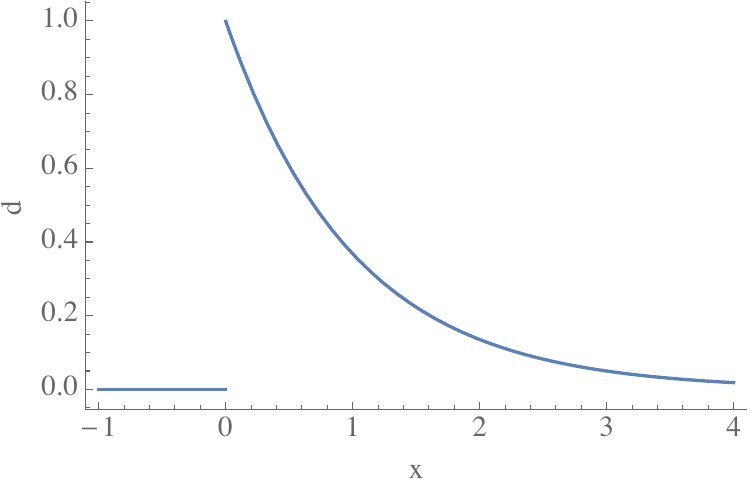}
        \put(60,50){\footnotesize{Prior}}
\end{overpic}
\begin{overpic}[width=0.45\textwidth]{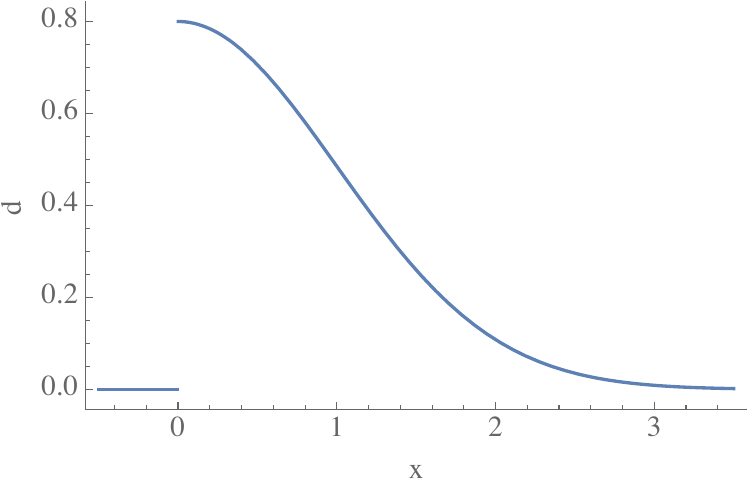}
        \put(60,50){\footnotesize{Posterior}}
\end{overpic}
        \caption{Prior and posterior model variable distributions for Example 3.} \label{fig:exponential}
\end{figure}

In this example, we look at a problem for which the prior distribution is not Gaussian, but for which a transformation to Gaussian variable can be introduced after which the previous methodology can be applied. The prior distribution for the model variable is exponential with mean and standard deviation equal to 1. A single measurement is made of ($g(x) = x + \epsilon$) with Gaussian observation error $\epsilon \sim N[0,0.36]$.   The observed value is $d\sbr{obs}= 1.$ The prior and posterior distributions for the model variable are shown in Fig.~\ref{fig:exponential}.
 
\begin{figure}[htbp!]
    \centering
\begin{overpic}[width=0.31\textwidth]{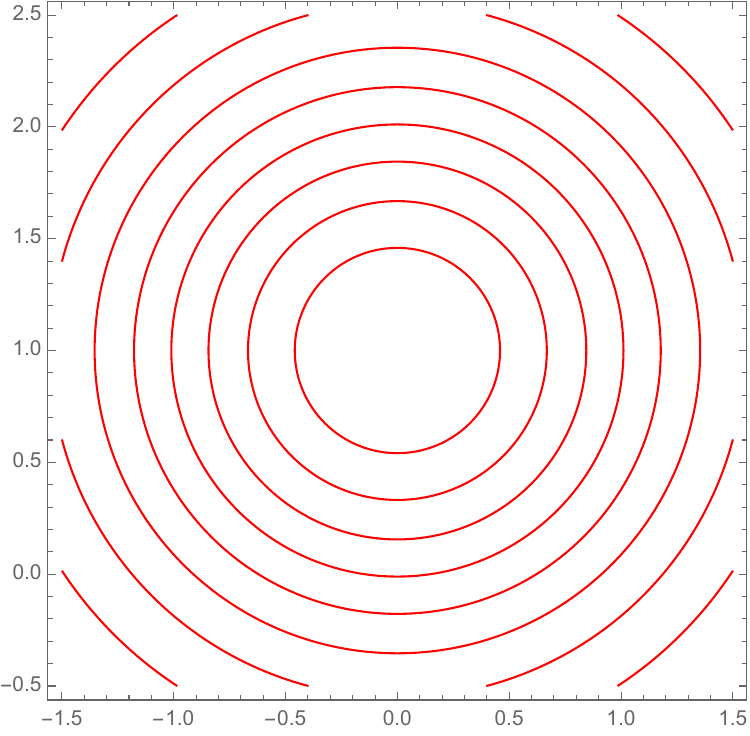}
        \put(10,85){\footnotesize{Prior density}}
\end{overpic}
\begin{overpic}[width=0.31\textwidth]{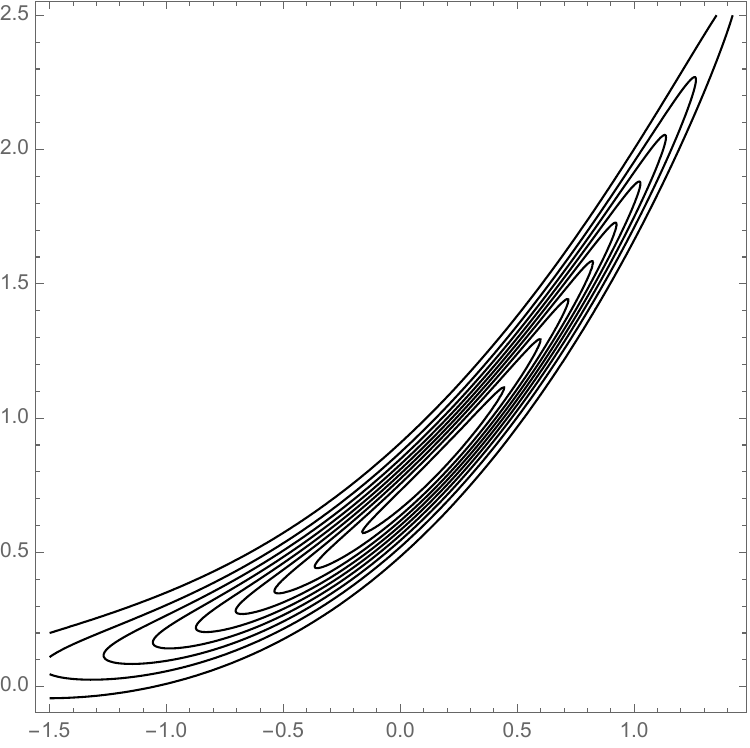}
        \put(10,85){\footnotesize{Posterior density}}
\end{overpic}
\begin{overpic}[width=0.31\textwidth]{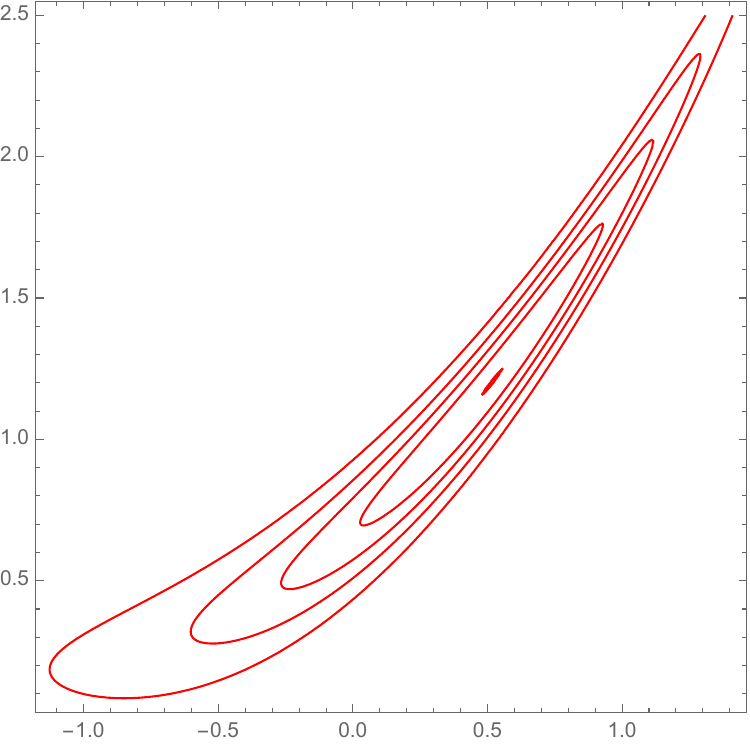}
        \put(10,85){\footnotesize{Proposal density}}
\end{overpic}
        \caption{Joint prior, posterior, and proposal distributions for Gaussian model variable and data variable. Model variable on horizontal axis.} \label{fig:exp_joint}
\end{figure}

Since the prior distribution in this example is not Gaussian, we define a new Gaussian variable $z$ related to $x$ as $z = F_z\inv \left( F_x (x) \right)$, where $F_x$ and $F_z$ are the cdfs for $x$ and $z$ respectively. After transformation of the model variable, the joint prior distribution for model and data, the joint posterior distribution, and the joint proposal distribution are shown in Fig.~\ref{fig:exp_joint}. The joint proposal density that results from minimization with $\rho = 0.25$ is   similar to the joint posterior density with $\gamma = 0.01$.

\begin{figure}[htbp!]
    \centering
    \begin{subfigure}[t]{0.31\textwidth}
        \centering
\includegraphics[width=0.95\textwidth]{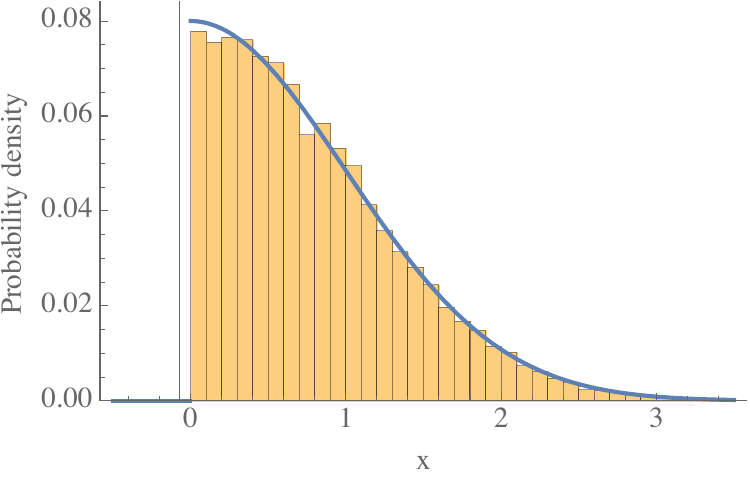}
        \caption{Distribution of samples from Algorithm 2 (histogram) compared with posterior distribution (solid curve).} \label{fig:exp_summary_a}
    \end{subfigure}%
    ~ 
    \begin{subfigure}[t]{0.31\textwidth}
        \centering
\includegraphics[width=0.95\textwidth]{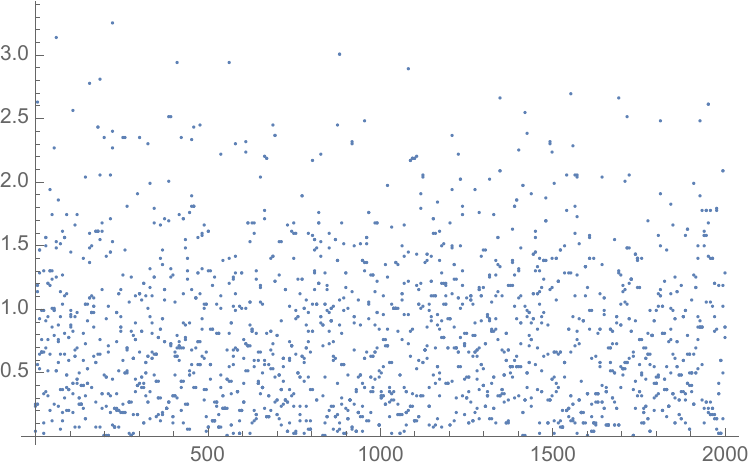}
        \caption{First 2000 elements of the Markov chain for the model variable.} \label{fig:exp_summary_b}
    \end{subfigure}%
    ~ 
    \begin{subfigure}[t]{0.31\textwidth}
        \centering
\includegraphics[width=0.95\textwidth]{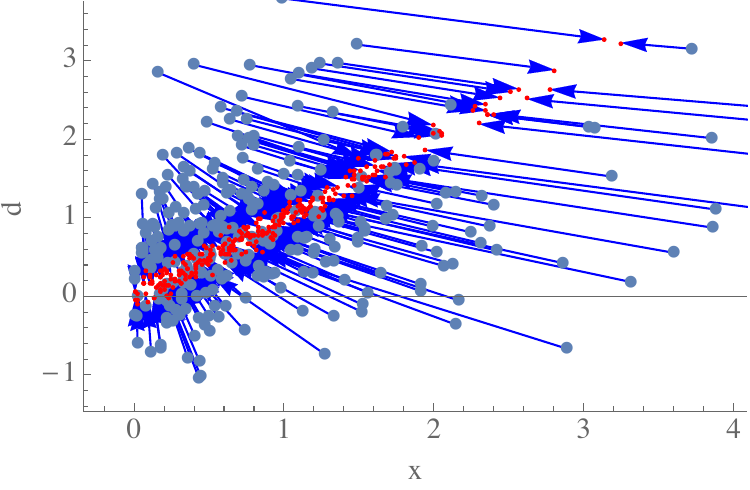}
        \caption{Mapping of variables from the prior to the posterior.} \label{fig:exp_summary_c}
    \end{subfigure}%
        \caption{Summary plots for Example 3 in the original variable with an exponential prior distribution.} \label{fig:exp_summary}
\end{figure}

An inverse transform must be applied to each sample from the Markov chain to obtain samples in the original model space. Fig.~\ref{fig:exp_summary_a} shows that the distribution of Monte Carlo samples is indistinguishable from the target posterior distribution. When this proposal density is used in the RML independence Meteropolis sampler (Algorithm \ref{alg:RML_new}) the acceptance rate is 74\%. 
Because the proposals are independent and the acceptance rate is high, the mixing in the chain is very good. Figure~\ref{fig:exp_summary_b} shows the first 2000 elements of the chain, with no evidence of a burn-in period or of nonstationary behavior.
Recall that the candidates are first generated from the joint prior, then a minimization is performed to place the proposals in regions of high probability.  Fig.~\ref{fig:exp_summary_c} shows the mapping of the first 200 accepted candidates in the Markov chain from the joint prior to the proposal distribution. The weighting has already been used for this figure so that the 26\% unsuccessful candidates are not shown.  

\section{Discussion}

In this paper, we proposed an augmented variable independence Metropolis sampler that uses minimization to place proposals in regions of high probability.   The computation required for a single proposal is considerably higher than the computation required for a typical MH algorithm, but because the proposals are independent and the acceptance rate is high, the cost per independent sample from the target distribution can be low compared to alternative methods.   The biggest limitation for application of the method to geoscience inverse problems will be the computation of the Jacobian determinant of the mapping that generates proposals. 


The augmented variable RML method described in this paper contains two parameters $\rho$ and $\gamma$ that must be tuned for efficiency and for validity of sampling. For linear observations, the proposal density and the target density can be made close by  setting both parameters to take small values. 
For multimodal posterior pdfs, however,  the proposal density $q(\vc{x}_\ast, \vc{d}_\ast) \not> 0$ for all $\pi(\vc{x}_\ast, \vc{d}_\ast) > 0$ for arbitrary choices of  $\rho$ and $\gamma$  so the Markov chain may not converge to $\pi(\cdot, \cdot)$ unless $\rho \ll 1$ and $\gamma \approx 1$. The acceptance rate for the numerical examples has relatively insensitive to the the value of $\gamma$ over a fairly large range.  Unlike typical applications of Metropolis random walk samplers, which often get stuck for long periods in a single mode of a multimodal distribution, the RML  Metropolis sampler is more likely to remain in the regions between modes as the proposal density tends to undersample from those regions. 

For cases in which the prior is not Gaussian, it can sometimes be possible to use anamorphosis to transform the variables so that the prior is Gaussian, in which cases the algorithm can be used for sampling. This is straightforward for single variable problems, but finding a transformation  to multivariate Gaussian in high dimensions seems impractical.

\section{Acknowledgements}

Primary support has been provided by the CIPR/IRIS cooperative research project ``4D Seismic History Matching'' which is funded by industry partners Eni, Petrobras, and Total, as well as the Research Council of Norway (PETROMAKS).

\bibliography{/Users/deanoliver/Documents/Bibliography/h_match/A-D,%
/Users/deanoliver/Documents/Bibliography/h_match/E-H,%
/Users/deanoliver/Documents/Bibliography/h_match/I-L,%
/Users/deanoliver/Documents/Bibliography/h_match/M-O,%
/Users/deanoliver/Documents/Bibliography/h_match/P-S,%
/Users/deanoliver/Documents/Bibliography/h_match/T-Z}

\bibliographystyle{siam}

\end{document}